\documentclass[aps,prl,10pt,twocolumn,floats,floatfix,,showkeys,showpacs,superscriptaddress]{revtex4}
\usepackage{graphicx,amssymb,amsmath}
\usepackage[latin1]{inputenc}
\usepackage{color}
\begin{document}

\title{Structure and function in flow networks}
\author{Nicol{\'a}s Rubido}
\affiliation{Institute for Complex Systems and Mathematical Biology, University of Aberdeen, King's College, AB24 3UE Aberdeen, UK}
\affiliation{Instituto de F\'{i}sica, Facultad de Ciencias, Universidad de la Rep\'{u}blica, Igu\'{a} 4225, Montevideo, 11200, Uruguay}
\author{Celso Grebogi}
\affiliation{Institute for Complex Systems and Mathematical Biology, University of Aberdeen, King's College, AB24 3UE Aberdeen, UK}
\author{Murilo S. Baptista}
\affiliation{Institute for Complex Systems and Mathematical Biology, University of Aberdeen, King's College, AB24 3UE Aberdeen, UK}
%
%
\date{\today}
\begin{abstract}
This Letter presents a unified approach for the fundamental relationship between structure and function in flow networks by solving analytically the voltages in a resistor network, transforming the network structure to an effective all-to-all topology, and then measuring the resultant flows. Moreover, it defines a way to study the structural resilience of the graph and to detect possible communities.
\end{abstract}

\maketitle
The relation between structure and function is one of the most studied topics in the theory of complex networks \cite{Girvan,Newman1,Newman2,Motter,Lai1,Lai2,Havlin}. The structure is a topological representation of the interacting elements forming a network and it is rigorously described by the theory of graphs. Behaviour is a functional observable of the network and can be measured by a variety of different approaches and methodologies, e.g., it can be quantified as a statistical description of the weights of connections of the structure of the network \cite{Girvan,Newman1,Newman2,Motter} or depend on the dynamics of interactions among the elements.

This work deals with flow networks that satisfy conservation laws (Kirchhoff's laws). The model for the flow network is stated in terms of resistor networks (weighted symmetric graphs), which have a source node $s$ and a sink node $t$ feeding the system, and a linear relationship between voltages and currents \cite{VF_problem}. The problem models a wide variety of physical phenomena, such as river flow channels, transport networks, ion channels, and electrical circuits. Its solutions establish a clear relationship between the topological structure of the networks (namely, adjacency matrix and edge weights, assumed known) and the functional flows passing through nodes and edges (that are a consequence of solving the flow model). The foundation of network flow theory roots back to Kirchhoff \cite{Kirchhoff}, answering what are the current flows in an electrical circuit when a set of voltages is applied. The solution is then achieved by solving Kirchhoff's equations \cite{Kirchhoff,Ahuja,Bollobas}.

In this Letter, we present a unified approach for the fundamental relationship between structure and function in flow networks by calculating voltages (loads) when input currents (flows) are given.
The main results of this approach are the following.

First, a novel formula for the voltages in the flow network model, i.e., the \emph{Voltage-Flow (VF) problem} [Eq.~(\ref{eq_V-F_problem})], is analytically derived. The solution [Eq.~(\ref{eq_voltages})] is expressed in terms of the known graph's weighted Laplacian matrix eigenspace base. In particular, we show that the behaviour of the flows is not only dependent on the matrix spectra, but intrinsically connected to the eigenvectors of this matrix.

Second, we show that the \emph{equivalent resistance} (also known as two-point distance resistance) formula \cite{Wu-Arpita} is retrieved [Eq.~(\ref{eq_resistance})] without further assumptions from the analytical expression for the voltages induced by the single $s$-$t$ pair. The equivalent resistance between any two nodes is a topological measure that gives further information about the structure of any graph. It is shown here how it constitutes a highly useful tool to reduce the complexity of any network structure: an arbitrary topology is transformed to an effective all-to-all complete graph.

Third, using both the voltage differences and the equivalent resistance, an \emph{effective functional flow adjacency} (EFFA) between any two points in the circuit is constructed [Eq.~(\ref{eq_eff_currents})]. These effective flows are the real measurable observables that are obtained when a probe is placed between any two nodes in the network. The EFFA matrix expresses the level of connectivity in terms of flow transmission; the functional relationship among the network components in the effective all-to-all topology. In particular, if two nodes are not directly connected, an effective current between them may still exist.

Fourth, as an application of our approach, we show how to identify the nodes that are mainly responsible for the transmission of flow, how to detect qualitatively communities of the network, and how to infer maximal node outflow for different graph topologies and resistor distributions. This reveals topological issues that are related to node maximum capacities in a natural manner.

The starting point for our unified approach is the calculation of voltages in a network by solving the VF problem. The VF problem is set by interpreting the graph structure as a linear resistor network circuit, which then receives an input flow $I$ at node $s$ and leaves at node $t$ (a current $I$ enters the circuit at a source and leaves at a sink). The network structure is given by a connected strict graph $\mathcal{G} = \{\mathcal{V},\,\mathcal{E}\}$ (where $\mathcal{V}$ and $\mathcal{E}$ are the sets of nodes and edges, respectively) with symmetric edge weights \mbox{$W_{ij} \equiv A_{ij}/R_{ij}$} for $i,\,j = 1,\ldots,\,N$ ($N$ being the number of nodes in $\mathcal{V}$, $A_{ij}$ the $ij$-th element of the adjacency matrix, and $R_{ij}$ the edge's resistance).

Imposing conservation of charge in every node of the network, i.e., a graph where at each node the current arriving at it equals the amount leaving it (first Kirchhoff's law), the flow vector is $(\vec{F}^{(st)})_i = I\,\left( \delta_{is} - \delta_{it} \right)$ for $i = 1,\,\ldots,\,N$, with $\delta_{ij}$ being the Kronecker delta. Thus, using the laws of Kirchhoff and Ohm, the net current in any node is given by
\begin{equation}
F_i^{(st)} = I\,\left(\delta_{is} - \delta_{it}\right) =  \sum_{j = 1}^N \frac{A_{ij} }{R_{ij}}\left( V_i^{(st)} - V_j^{(st)} \right)\,,
 \label{eq_init_flows}
\end{equation}
where $V_i^{(st)}$ is the voltage potential at node $i$ given the particular $s$-$t$ pair. Equation (\ref{eq_init_flows}) is rearranged such that the VF problem is expressed in matrix form
\begin{equation}
  \mathbf{G}\,\vec{V}^{(st)}  = \vec{F}^{(st)}\,,
 \label{eq_V-F_problem}
\end{equation}
where the upper-indexes indicate that the problem depends on the location of the $s$-$t$ nodes on $\mathcal{V}$ and $\mathbf{G}$ is the weighted Laplacian matrix. The entries of $\mathbf{G}$ are
\begin{equation}
 G_{ij} = \left\lbrace \begin{array}{lcl}
                         \sum_{k=1}^N W_{ik} & \text{if} & i = j\,, \\
			  - W_{ij} & \text{if} & i \neq j\,.
                        \end{array} \right.
 \label{eq_laplacian_def}
\end{equation}
Then, the VF solution is achieved once the voltages in each node are found from inverting $\mathbf{G}$. However, the rank of $\mathbf{G}$ is $N-1$ ($\det\left( \mathbf{G} \right) = \prod_{n = 0}^{N-1}\lambda_n = 0$) and direct inversion is not possible \cite{Bollobas,FanChung}.

The Laplacian inversion is addressed here by a different interpretation of the derivation of the generalized inverse matrix, also known as Moore-Penrose matrix \cite{Bollobas,Wu-Arpita}.

The eigenvalue problem for the Laplacian is given by $ \mathbf{G}\,\mathbf{P} = \mathbf{P}\,\mathbf{\Lambda}$, where $\Lambda_{ij} = \delta_{ij}\,\lambda_{n}$ ($n = i-1 = 0,\ldots,N-1$) is a diagonal matrix containing all eigenvalues and \mbox{$\mathbf{P} = \{\vec{v}_0,\,\vec{v}_1,\,\ldots,\,\vec{v}_{N-1}\}$} is the matrix whose columns are the eigenvectors of $\mathbf{G}$. We prove here that to express any element of $\mathbf{G}$, the only elements needed from the spectral decomposition are the non-null eigenvalues and corresponding eigenvectors. This results in the following decomposition of $\mathbf{G}$
\begin{equation}
  \mathbf{G} = \mathbf{P}_r\,\mathbf{\Lambda}_r\,\mathbf{P}_r^{T}\,,
 \label{eq_eigen_decomp}
\end{equation}
where $T$ indicates transpose, $\mathbf{P}_r \in \mathbb{R}^{N\times(N-r)}$, $\mathbf{\Lambda}_r \in \mathbb{R}^{(N-r)\times(N-r)}$, and $r$ being the number of connected components of $\mathcal{G}$, which in this Letter is fixed at $r = 1$.

Equation (\ref{eq_eigen_decomp}) is derived by analyzing each Laplacian matrix entry using the full spectral decomposition. In particular, $G_{ij} = \sum_{k=1}^N P_{ik} \lambda_{k-1} P^{-1}_{kj}$, where for $k = 1$, $\lambda_0 = 0$, so no contribution is added to the sum. Thus, observing that the inverse of an orthogonal matrix is its transpose ($P^{-1}_{kj} = P_{jk} = \left(\vec{v}_{k-1}\right)_j$), the $ij$'s entry of $\mathbf{G}$ is given by: $G_{ij} = \sum_{k=2}^N \left(\vec{v}_{k-1}\right)_i\, \lambda_{k-1}\,\left(\vec{v}_{k-1}\right)_j$, which is easily extendible to various connected components making the lower bound for the summation index $k = r+1$. 

However, removing the null eigenspace of the spanning set of eigenvectors, the rank of the new $\mathbf{G}$ in Eq.~(\ref{eq_eigen_decomp}) is $N$ and inversion can be directly performed. Consequently, each element of the generalized inverse Laplacian matrix $Z_{ij} \equiv (\mathbf{G}^{-1})_{ij}$ is found from
\begin{equation}
  Z_{ij} = \sum_{n = 1}^{N-1} \left(\vec{v}_n\right)_i \frac{1}{\lambda_n} \left(\vec{v}_n\right)_j\,,
 \label{eq_laplacian_inv}
\end{equation}
with the new index $n = k-1$ and $i,\,j = 1,\ldots,N$. It is easy to show that $\sum_{k=1}^N G_{ik}\,Z_{kj} = \delta_{ij} - 1/N$, where the extra constant factor comes from the incompleteness of the restricted eigenspace and it cancels out when calculating voltage differences.

Returning to Eq.~(\ref{eq_V-F_problem}), the \emph{solution for the VF problem} at a given node $i\in\mathcal{V}$ is $ \left( \vec{V}^{(st)} \right)_i = \sum_{j = 1}^N Z_{ij}\,\left( \vec{F}^{(st)} \right)_j = I\,\left( Z_{is} - Z_{it} \right)$,
\begin{equation}
  \left( \vec{V}^{(st)} \right)_i = I\,\sum_{n = 1}^{N-1} \frac{ \left(\vec{v}_n\right)_i }{ \lambda_n } \left[\left( \vec{v}_n \right)_s -  \left( \vec{v}_n \right)_t\right]\,.
 \label{eq_voltages}
\end{equation}
This formula is the backbone of our unified approach. It is applicable to any connected weighted graph and is extendible to many sources and sinks with different inputs and outputs as long as flow conservation is fulfilled ($\sum_i F_i^{(st)} = 0$ and $F_i = 0\;\forall\,i\neq s$ or $t$). Furthermore, it allows to derive the equation for the equivalent resistance between any two nodes, to construct the effective flows, and later apply them to, e.g., community detection.

The derivation of the \emph{equivalent resistance} $\rho_{ij}$ between any two nodes \cite{Wu-Arpita} is done as follows. Set the source at a starting node ($i = s$) and the sink at an ending point ($j = t$), then Eq.~(\ref{eq_voltages}) provides an exact closed formula for $\rho_{ij}$ in terms of eigenvalues and eigenvectors of the weighted Laplacian matrix. The reason for doing this is that the voltage difference between these two nodes gives the incoming current times the resistance between them, hence, $(\vec{V}^{(st)})_s - (\vec{V}^{(st)})_t = I\,\rho_{st}$. Therefore, using the same procedure for every pair of nodes in the graph all $\rho_{ij}$'s are determined,
\begin{equation} 
 \rho_{ij} = \sum_{n = 1}^{N-1} \frac{ 1 }{ \lambda_n } \left[\left( \vec{v}_n \right)_i - \left( \vec{v}_n \right)_j\right]^2\,,
 \label{eq_resistance}
\end{equation}
providing a solution that is independent of where the $s$-$t$ pair is placed.

The matrix element $\rho_{ij}$ is a topological measure that weighs all paths between any pair of nodes $i$ and $j$. All paths between these nodes that are allowed by the adjacency structure of the graph are collapsed to one equivalent link with weight $\rho_{ij}$. The result is that a single effective link is created between $i$ and $j$. The important contribution is that \emph{this quantity presents the possibility to diminish the complexity of any graph to a simpler all-to-all equivalent topology} with a betweenness score given by $\rho_{ij}$.

\begin{figure}[htbp]
 \begin{center}
  \includegraphics[width=0.48\columnwidth]{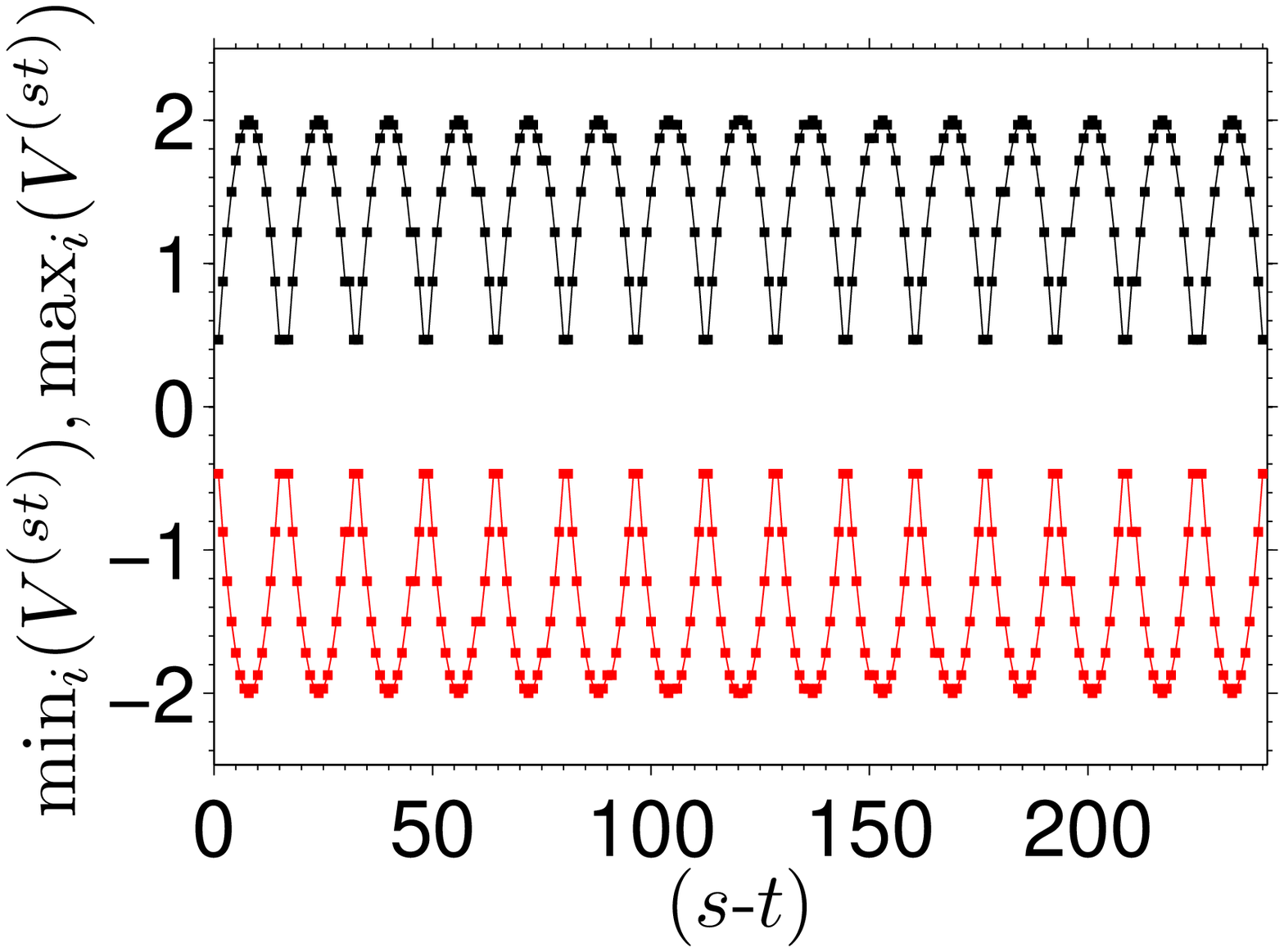}
  \includegraphics[width=0.47\columnwidth]{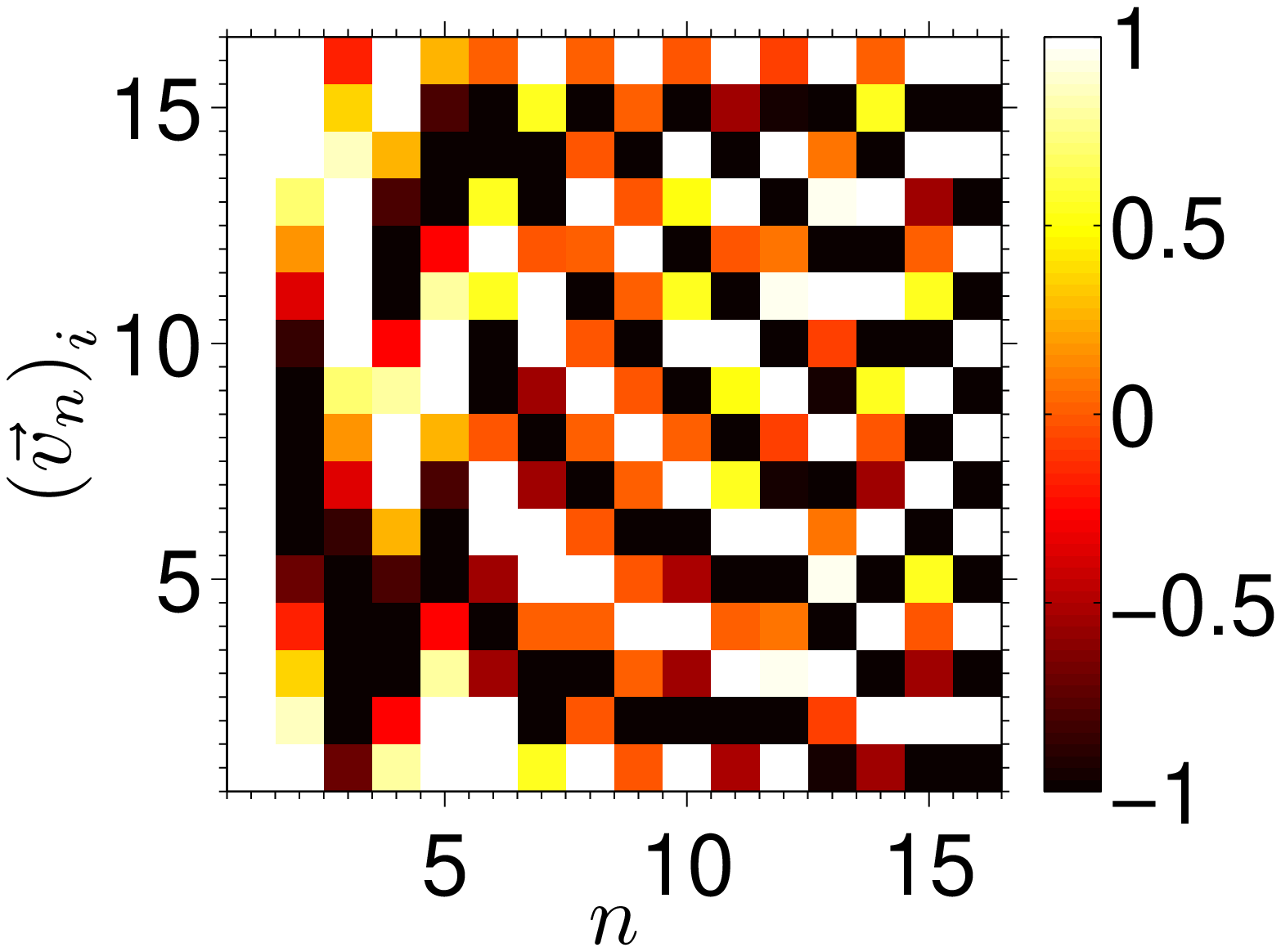} \\
  \includegraphics[width=0.48\columnwidth]{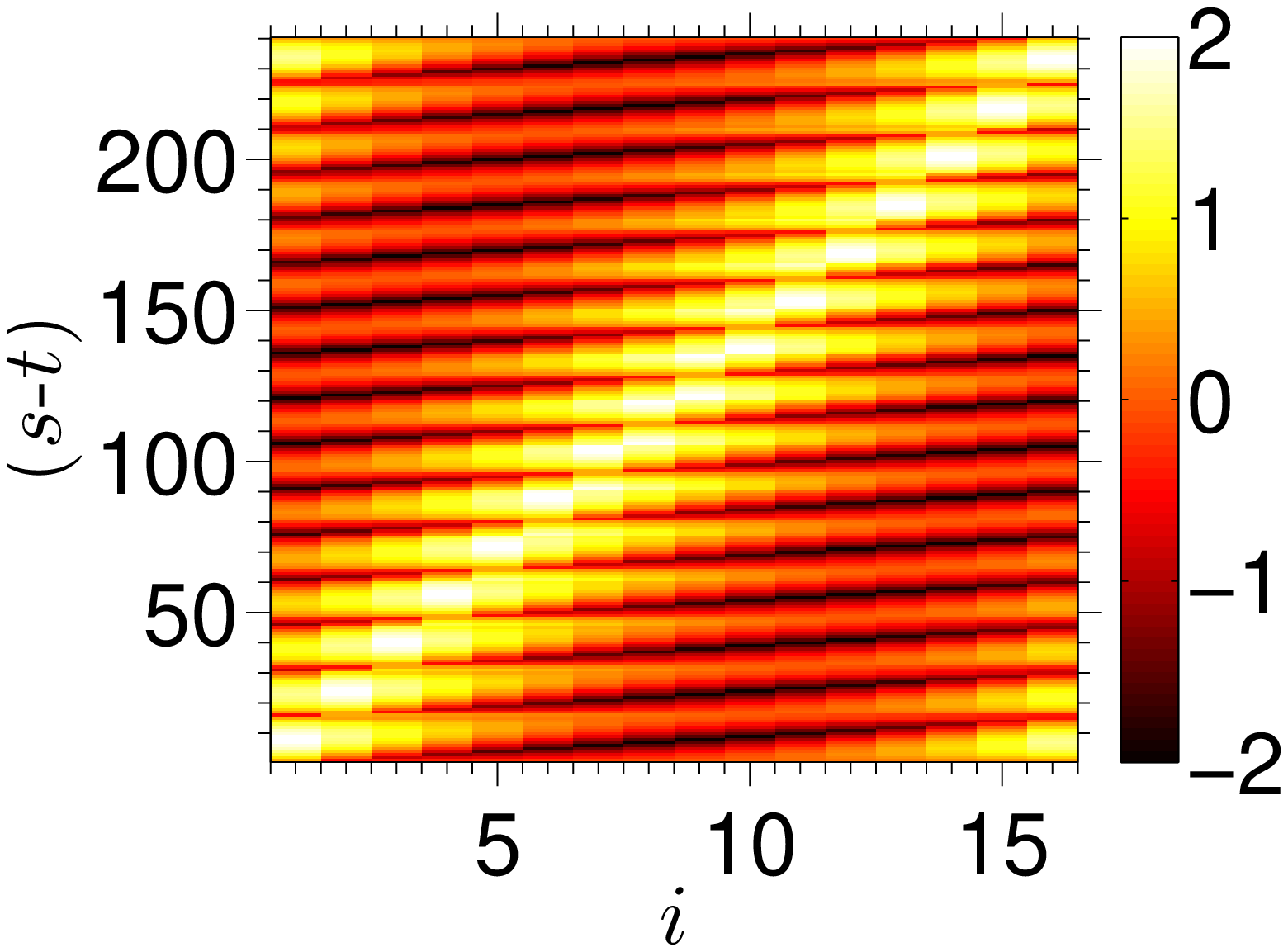}
  \includegraphics[width=0.48\columnwidth]{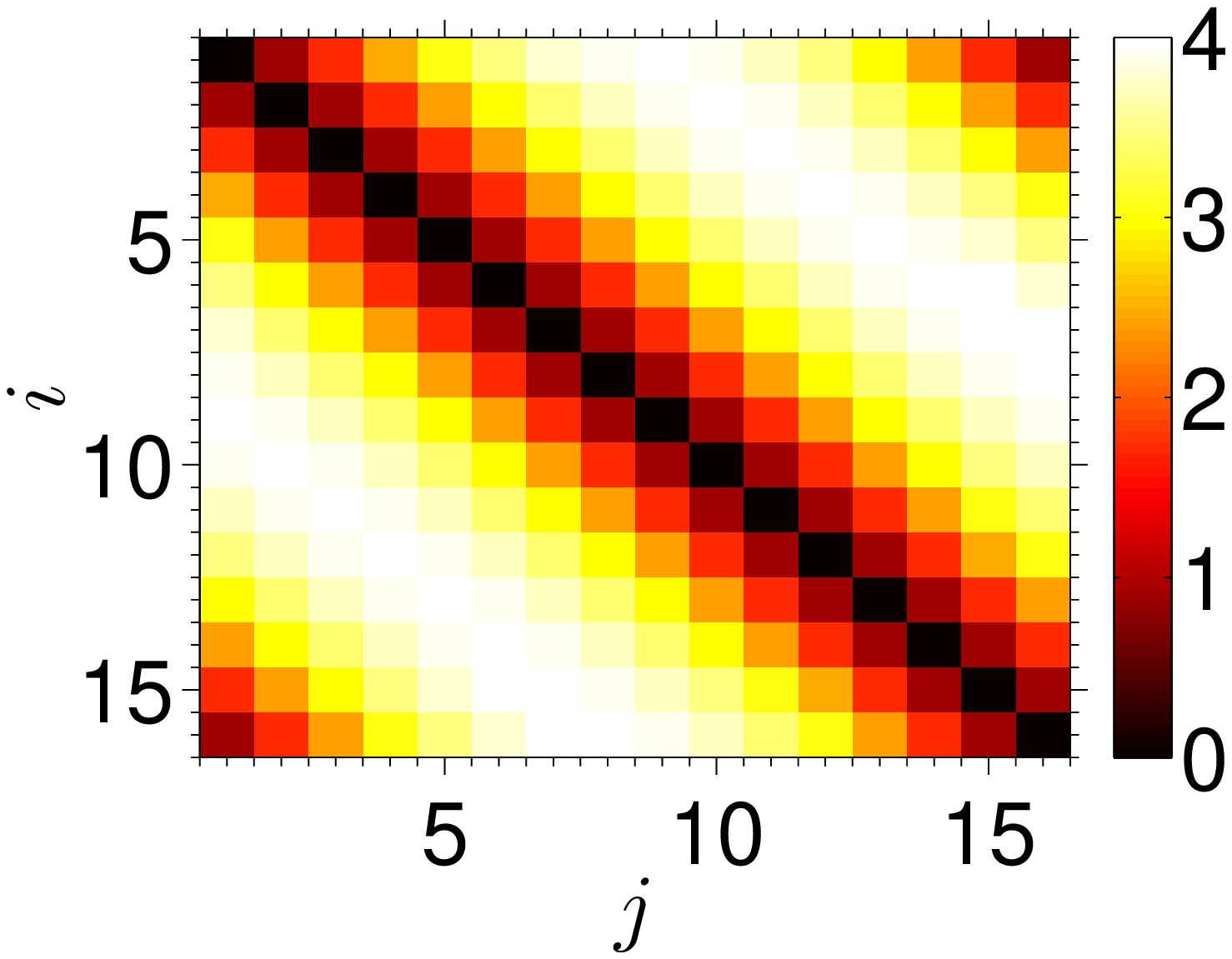}
 \end{center} \vspace{-1.0pc}
  \caption{The top left panel has the maximum (black) and minimum (red) voltages that are achieved in an unweighed ($R_{ij} = 1\;\forall\,i,j$) ring graph ($C_N$) with $N = 16$ nodes for every $s$-$t$ pair in units of $I = 1$ for the input current at node $s$. The top right panel shows the eigenvectors ($\sqrt{N} \vec{v}_n$) coordinate values (color scale) of the corresponding weighed Laplacian matrix. The bottom left panel exhibits in color coded, for all these possible input-output configurations, the voltages $V_i^{(st)}$ of every node. The bottom right panel shows the equivalent resistor matrix ($\rho_{ij}$) for every pair of nodes in $C_N$.}
 \label{fig_CN_VFproblem}
\end{figure}

As a working example, results of applying the solution given by Eqs.~(\ref{eq_voltages}) and (\ref{eq_resistance}) to a ring graph, namely $C_N$, of $N = 16$ nodes with resistors equal to one ($R_{ij} = 1\;\forall\,i,j$) and input current $I = 1$ are shown in Fig.~\ref{fig_CN_VFproblem}. For this network, the set of unordered eigenvalues is given by \mbox{$\lambda_n = 2 - 2\cos(2\pi\,n/N)$}, with $n = 0,\ldots,N-1$ \cite{FanChung}.

The top left panel in Fig.~\ref{fig_CN_VFproblem} shows that for the $N\,(N-1) = 240$ possibilities of $s$-$t$ pairs in $C_N$ the maximum voltages achieved in the ring do not surpass $2$ (arbitrary units). This is the maximum difference that the eigenvector coordinates can achieve in the non-normalized frame (top right panel in Fig.~\ref{fig_CN_VFproblem}), hence giving a maximum voltage difference of $4$. This result shows that not only the eigenvalues of the Laplacian are important to infer the network's functional behaviour, but eigenvectors are also needed. Also, the maximum (minimum) is periodic, in the sense that the largest difference happens every time the source and sink are separated by $8$ edges.

The bottom panels in Fig.~\ref{fig_CN_VFproblem} show the voltage solutions for all the $s$-$t$ pairs (left) and all equivalent resistances (right). The resulting values show the functional and structural symmetry that this simple network has. As all edge weights are unity ($R_{ij} = 1$), the equivalent resistance $\rho_{ij}$ maximum can only be $4$. This is easily seen when calculating it directly by means of series and parallel resistor calculations. For instance, for nodes $i = 1$ and $j = 9$ there are two parallel paths with $8$ edges in series, hence: $ \rho_{1,9}^{-1} = \frac{1}{8} + \frac{1}{8} \;\Rightarrow\; \rho_{1,9} = 4$, which is the same value as element $\rho_{1,9}$ in the bottom right panel of Fig.~\ref{fig_CN_VFproblem} (found from Eq.~(\ref{eq_resistance})).

Having the voltage difference and equivalent resistance given by Eqs.~(\ref{eq_voltages}) and (\ref{eq_resistance}), the third result of our unified approach is addressed: the construction of the \emph{effective edge flows}, namely, $\phi_{ij}^{(st)}$. We introduce these flows by finding the voltage differences among any two nodes in the effective topology given by $\rho_{ij}$. Thus, the EFFA matrix elements for each $s$-$t$ configuration are given by
\begin{equation}
  \phi_{ij}^{(st)} = \frac{ 1 }{ \rho_{ij} } \left[ \left( \vec{V}^{(st)} \right)_i - \left( \vec{V}^{(st)} \right)_j \right]\,,
 \label{eq_eff_currents}
\end{equation}
where $(\vec{V}^{(st)})_i > (\vec{V}^{(st)})_j$ such that an effective current is flowing from node $i$ to $j$, otherwise $\phi_{ij}^{(st)} = 0$. Hence, the $s$-$t$ dependent EFFA matrix defines an effective directed network, which breaks the symmetry of the effective all-to-all topology structure given by $\rho_{ij}$. In particular, the physical current on each edge $ij\in\mathcal{E}$ of the graph is given by $I_{ij}^{(st)} = \frac{ A_{ij} }{ R_{ij} }\,[ ( \vec{V}^{(st)} )_i - ( \vec{V}^{(st)} )_j ]$.

The voltage differences define a functional relationship among nodes and the equivalent resistance an effective topology structure. The introduction of \emph{the EFFA relates both structure and functional characteristics of the graph}, thus, it enables to find functional hubs that might not be there in the topological structure, and detect highly connected areas (communities) in a functional way. In this sense, the $s$-$t$ analysis acts similarly as how a dye does when introduced to a circulatory system of a person: it highlights a certain part of the network structure to detect what is being sought, e.g., clog blocking, arteries stiffness, etc.

\begin{figure}[htbp]
 \begin{center}
  \includegraphics[width=0.48\columnwidth]{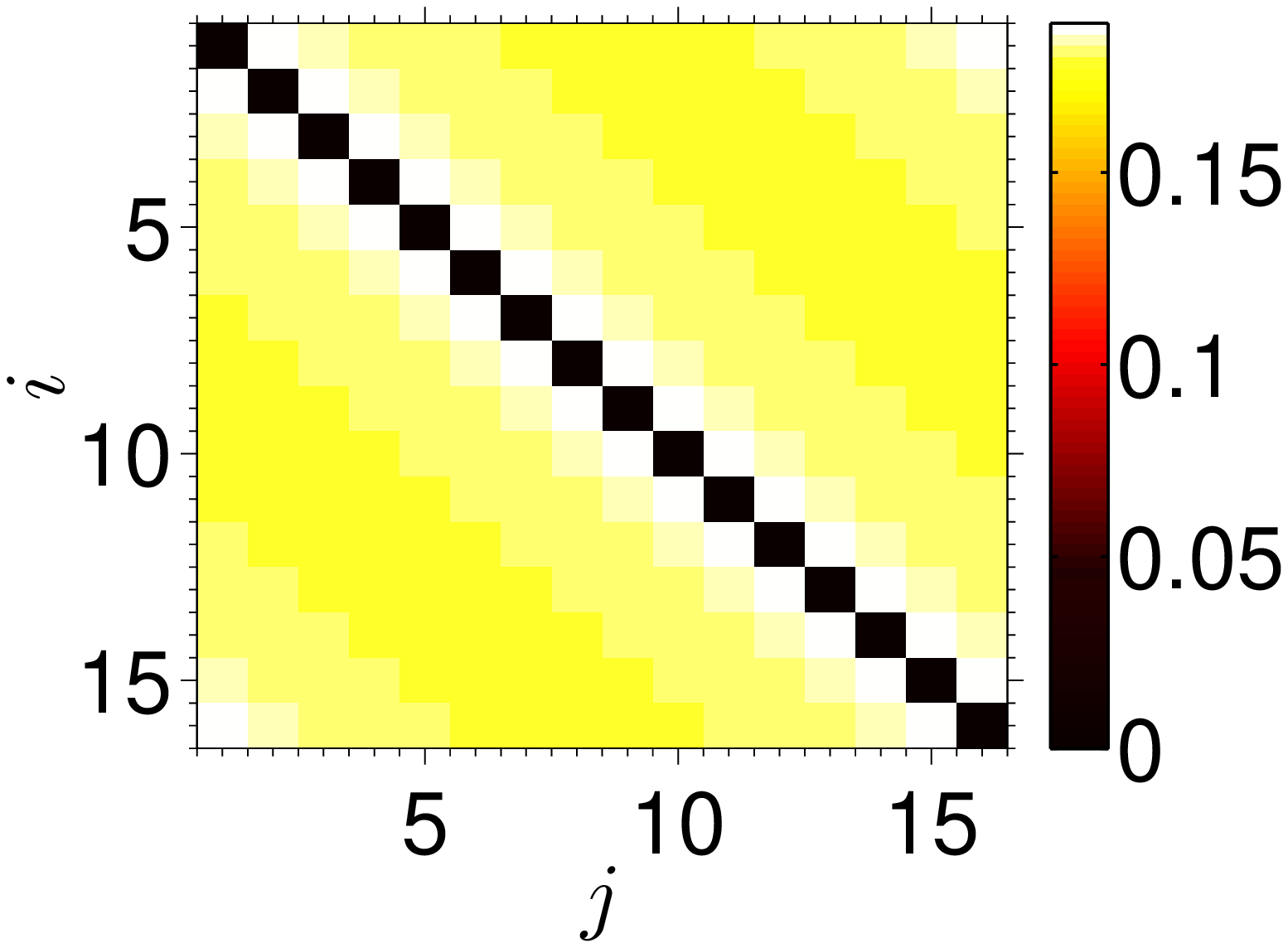}
  \includegraphics[width=0.48\columnwidth]{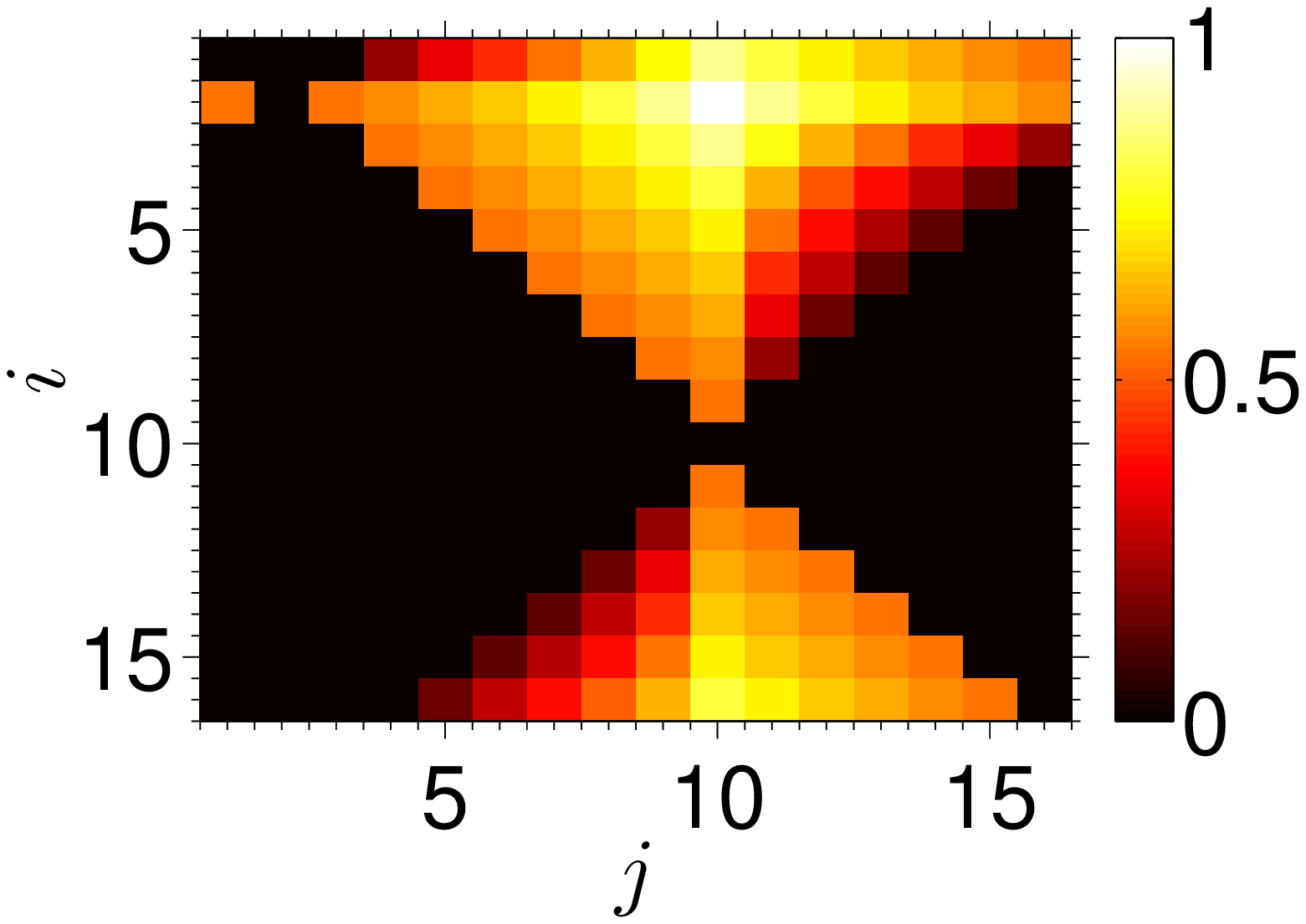}
 \end{center} \vspace{-1.0pc}
  \caption{The left panel shows the effective flow functional adjacency (EFFA) matrix for the unweighed $C_N$ graph of Fig.~\ref{fig_CN_VFproblem}, where all source-sink pairs have been averaged, i.e., $\phi_{ij}$. The right panel exhibits the same EFFA matrix for a particular source-sink pair, namely, $s = 2$ and $t = 10$ ($\phi_{ij}^{(2,10)}$). Each matrix element's magnitude is associated with a color scale.}
 \label{fig_eff_C_N}
\end{figure}

The mean EFFA matrix is defined to detect statistically relevant functional observables. It is calculated by all the effective currents (Eq.~(\ref{eq_eff_currents})) among the different pairs averaged over all possible realizations of source-sink pairs. Its matrix elements are
\begin{equation}
  \phi_{ij} \equiv \left\langle \phi_{ij}^{(st)} \right\rangle_{(st)} = \sum_{s = 1}^N \sum_{t = 1\,(t \neq s)}^N \frac{ \phi_{ij}^{(st)} }{N\,(N-1)}\,,
 \label{eq_EFFA_entries}
\end{equation}
which define a matrix that constitutes an effective structural flow quantity and is related to the equivalent resistance matrix. The EFFA matrix and its averaged counterpart for the example of the ring graph are shown in Fig.~\ref{fig_eff_C_N}, where the left panel shows the mean EFFA matrix and the right panel exhibits the symmetry breaking for a particular $s$-$t$ pair.

\begin{figure}[htbp]
 \begin{center}
    \includegraphics[width=0.9\columnwidth]{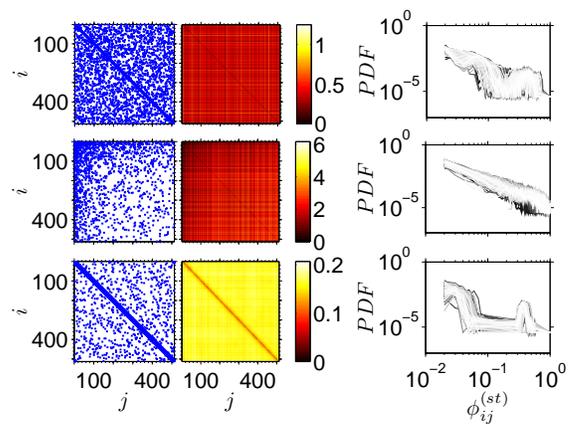}
 \end{center} \vspace{-1.5pc}
  \caption{The figure shows the following unweighted graphs: a random (top row), a scale-free (centre row), and a small-world (bottom row) network of $N = 2^9$ nodes and connecting probability $p = 0.05$ \cite{graphs}. The adjacency matrix (left column), equivalent resistance matrix (centre column in color coded), and effective functional flow adjacency PDFs (right column) for all $s$-$t$ configurations (gray scale) are shown for each graph.}
 \label{fig_PDFs_EFFA}
\end{figure}

\begin{figure}[htbp]
 \begin{center}
  \includegraphics[width=0.9\columnwidth]{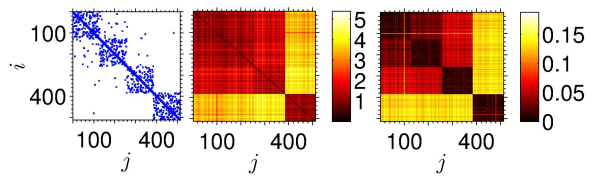}
 \end{center}\vspace{-1.5pc}
  \caption{The figure shows the adjacency matrix (left column), equivalent resistance (middle column), and the average effective functional adjacency matrix (right column) for an unweighed clustered network formed by $4$ random graphs of $N_c = 2^7$ \cite{graphs}. The network has a total of $N = 2^9$ nodes and each matrix element magnitude is associated with a color scale.}
 \label{fig_EFFAs}
\end{figure}

With the three quantities analytically expressed (voltages in Eq.~(\ref{eq_voltages}), equivalent resistances in Eq.~(\ref{eq_resistance}), and the effective currents in Eq.~(\ref{eq_eff_currents})), we investigate \cite{graphs} random graphs (RN) \cite{Erdos}, small-world (SW) networks \cite{Strogatz}, scale-free (SF) topologies \cite{Barabasi}, and numerically generated clustered networks (CN) of $N = 2^9$ nodes. The following conclusions constitute the application of our unified approach to these concrete topologies. In particular, the differences between the respective adjacency and effective topologies (equivalent resistance) are shown in the left and centre columns of Figs.~\ref{fig_PDFs_EFFA} and \ref{fig_EFFAs}.

To study other features of the EFFA matrix that are not represented in its averaged version and to answer how the flow is effectively distributed in any given topology, the probability density function (PDF) of the effective edge flows $\phi_{ij}^{(st)}$ for every particular $s$-$t$ is calculated. This does not only tells how diffusive or localized  the transport of currents is as a function of the effective topology, but also gives the maximum flow values a particular graph can have, i.e., the maximum edge capacity.

Results show that the EFFA PDFs for RN, SW, and SF networks are roughly invariant under changes of the \mbox{$s$-$t$} pair. However, SW and SF networks produce more high flow magnitudes ($\phi_{ij}^{(st)} \simeq 1$) than RN for a broad range of parameters \cite{graphs}, hence, RNs distribute flows in a more uniform way than SF or SW networks. These observations are seen in the example shown on the right column of Fig.~\ref{fig_PDFs_EFFA}.

The detection of communities is done by analysis of the mean EFFA matrix and the equivalent resistance. In general, \emph{the application of the EFFA analysis on numerically generated CNs qualitatively exhibits the communities that the networks have}. This is seen in the centre and right panels of the example exhibited in Fig.~\ref{fig_EFFAs}, where the communities are spotted as dark areas, both in $\rho_{ij}$ and $\phi_{ij}$, though the contrast in the mean EFFA matrix is higher than in the equivalent resistance, implying that communities are better identified by $\phi_{ij}$. \emph{The $\phi_{ij}$ matrix} not only detects the communities but also \emph{shows that the effective flows between communities are larger than the ones within each community}. Hence, the intra-edges connecting communities must have a high flow capacity as they represent the vulnerable links in the network. Removal of any of these edges generate drastic changes in the flows, leading to the isolation of communities.

The observations on flow transmission and community detection have been previously shown by doing structural analysis of networks, e.g., in Refs.~\cite{Girvan,Newman1,Newman2,Motter,Lai1,Lai2,Havlin}. In particular, the possibility of using voltages as a betweenness measure score is pointed out in Refs.~\cite{Girvan,Newman1,Newman2} and is related to shortest-paths and random walks. With our unified approach, and the four results that are addressed in this Letter, we show the practicality of it. Moreover, the linear relationship between loads and flows, and conservation of charge, are restrictions needed to define the flow network on top of the structure but not for the structure itself. Thus, the flow analysis is applicable to any network structure outside the domain of flow networks as long as the graph to be analysed is connected.

An estimate of the computational time that our approach requires is as follows. To calculate the voltages one needs the eigenvectors and eigenvalues of the Laplacian matrix. This is calculated in time $\mathcal{O}(N^{\alpha})$, with $\alpha$ being an exponent that nowadays rounds $2$ and $N$ the number of nodes in the network. Then, if all possible source-sink pairs are analysed, the flows require $\mathcal{O}(N^3)$ to be found. This means that the algorithm is not as effective as other methods \cite{Andrea} but its mathematical formulas still provide great information for obtaining upper and lower bounds for currents without the need for simulations. Furthermore, community detection can be quantitatively described if other techniques are used, such as PDF analysis of the EFFA matrix elements.


Authors acknowledge the Scottish University Physics Alliance (SUPA).




\begin{thebibliography}{10}
\bibitem{VF_problem} In particular, the approach that we take is derived from the maximum flow problem, which corresponds to finding a feasible flow through a single-source $s$ and a single-sink $t$ that is maximum when costs and capacities at the edges are given. The solution for this problem is equal to the minimum capacity of an $s$-$t$ cut in the network as stated in the max-flow min-cut theorem \cite{Ahuja,Bollobas}. On the other hand, if the $s$-$t$ flow is fixed, then the problem turns into finding how is the transmission given and what is its implications to the structure and function of the network.
\bibitem{graphs} Starting with a ring graph of $N = 2^9$ nodes, a tuning parameter $p$ controls the addition of links. For the RN, SW, and SF graphs the process is straightforward (taken from Refs.~\cite{Erdos,Strogatz,Barabasi}) and $p$ is related to the probability of adding an edge.
 The RN process follows the strategy of \cite{Erdos}. The SF graph is constructed as in the rewiring procedure of Ref.~\cite{Barabasi}.
 The SW graph is constructed as in Ref.~\cite{Strogatz}. 
 For the CN, four RN of $N_c = 2^7$ are created with equal statistics (probability $p = 0.01$), then approximately $10$ links are added between them randomly. In addition, weights are included from the uniform random distribution such that $R_{ij}\in(0,1]\; \forall\, i,j$.
\end{thebibliography}
\end{document}